\documentclass[12pt,a4paper]{article}

\usepackage{graphicx}
\usepackage{wrapfig}
\usepackage{epsfig}

\begin{document}
\textwidth=135mm
 \textheight=200mm

\begin{center}
{\bfseries Importance of fragmentation functions in determining
polarized parton densities \footnote{{\small A talk given at the
20th International Symposium on Spin Physics (SPIN2012) JINR,
Dubna, Russia, September 17 - 22, 2012. }}} \vskip 5mm E.
Leader$^\dag$, A. V. Sidorov$^\ddag$ and D. B. Stamenov$^{*}$
\vskip 5mm
{\small {\it $^\dag$ Imperial College, Prince Consort Road, London SW7 2BW, England}} \\
{\small {\it $^\ddag$ Joint Institute for Nuclear Research, 141980 Dubna, Russia}} \\
{\small {\it $^{*}$ Institute for Nuclear Research and Nuclear
Energy, Bulgarian Academy of Sciences, Blvd. Tsarigradsko Chaussee
72, Sofia 1784, Bulgaria}}
\\

\end{center}
\vskip 5mm \centerline{\bf Abstract} New fragmentation functions
(FFs) are extracted from a NLO QCD fit to the {\it preliminary}
COMPASS data on pion and kaon multiplicities. It is shown that the
new kaon FFs are very different from those of De Florian at al.
(DSS) and Hirai et al. (HKNS). The sensitivity of the extracted
polarized PDFs to the new FFs is discussed.

\vskip 10mm

In the absence of charged current neutrino data, the experiments
on polarized inclusive deep inelastic lepton-nucleon scattering
(DIS) yield information only on the sum of quark and antiquark
parton densities $\Delta q + \Delta \bar{q}$ and the polarized
gluon density $\Delta G$. In order to extract separately $\Delta
q$ and $\Delta \bar{q}$ other reactions are needed. One
possibility is to use the {\it polarized} semi-inclusive
lepton-nucleon processes (SIDIS) $l+ N \rightarrow l'+h+X$, where
$h$ is a detected hadron (pion, kaon, etc) in the final state. In
these processes new physical quantities appear - the fragmentation
functions $D^h_{q, \bar q}(z, Q^2)$ which describe the
fragmentation of quarks and anti-quarks into hadrons. Due to the
different fragmentation of quarks and anti-quarks, the polarized
parton densities $\Delta q$ and $\Delta \bar{q}$ can be determined
separately from a combined QCD analysis of the data on inclusive
and semi-inclusive asymmetries. However, for their correct
determination a good knowledge of the fragmentation functions is
very important. It turned out that the use of the DSS set of FFs
\cite{DSS} leads to the so called strange quark polarization
puzzle \cite{deltas_puzzle}, {\it i.e.} the contradiction between
the negative polarized strange quark density obtained from
analyses of inclusive DIS data alone and the positive values for
this density, for most of the range of measured {\it x}, obtained
from combined analyses of inclusive and semi-inclusive SIDIS data.
The significant difference in the kaon sector between the DSS FFs
and the other sets of FFs \cite{other_FFs} results from the use of
the {\it unpublished} HERMES'05 unpolarized SIDIS data on the
hadron multiplicities, used only in the DSS analysis. However, the
new preliminary HERMES data \cite{HERMES}, as well as the COMPASS
ones \cite{COMPASS} are not in agreement with the DSS FFs and
therefore, this set can not be favored at present. It is important
to mention that the data on hadron multiplicities in the
unpolarized SIDIS processes are crucial for a reliable
determination of FFs, because only they can separate
$D_q^h(z,Q^2)$ from $D_{\bar q}^h(z,Q^2)$.

In this talk we present our results on new fragmentation functions
extracted from the {\it preliminary} COMPASS 2004 deuteron data on
pion and kaon multiplicities \cite{COMPASS} in NLO QCD
approximation. We discuss also their impact on the determination
of the polarized sea quark densities, and in particular, the
status of the strange quark polarization puzzle.

The multiplicitiy $M_d^h(x,Q^2,z)$ of hadrons of type $\it h$
$(h=\pi^{+}, \pi^{-}, K^{+}, K^{-})$ using a deuteron target is
defined as the number of hadrons produced, normalized to the
number of DIS events, and can be expressed in terms of the
semi-inclusive cross section $\sigma_d^h$ and the inclusive cross
section $\sigma_d^{DIS}$:
\begin{equation}
M_d^h(x,Q^2,z)=
\frac{d^3N^h(x,Q^2,z)/dxdQ^2dz}{d^2N^{DIS}(x,Q^2)/dxdQ^2}
=\frac{d^3\sigma_d^h(x,Q^2,z)/dxdQ^2dz}{d^2\sigma_d^{DIS}(x,Q^2)/dxdQ^2}.
\label{M_exp}
\end{equation}

The data on the multiplicities are presented in different ways. In
our fit we have used the $2D(x, z)$ presentation. In this
presentation the values of the multiplicities are given for four
z-bins [0.2-0.3; 0.3-0.45; 0.45-0.65; 0.65-0.85] as a function of
different $(x, Q^2)$ bins. Note that for a given z-bin and given
$(x_i, Q^2_i)$-bin the multiplicity corresponds to the {\it
average} number of hadrons detected.

We will discuss here mainly our results on the fit to the data on
kaon multiplicities for two reasons: {\it i)} the big difference
between the kaon FFs obtained by the different groups, and {\it
ii)} the set of kaon FFs used in the combined analysis of the
polarized inclusive and semi-inclusive DIS data is crucial for the
determination of the polarized strange density. In the analysis 90
experimental points (45 for $K^{+}$ and 45 for $K^{-}$) have been
taken. The errors used are quadratic combinations of the
statistical error and half of the systematic error due to the kaon
identification by the RICH detector. The number of free
parameters, attached to the input parametrizations of the kaon FFs
[$D_u^{K+}(z),~D_{\bar u}^{K+}(z),~D_{\bar s}^{K+}(z),
D_g^{K+}(z)$] at $Q^2= 1~GeV^2$ and determined from the fit, is
13. The assumption that all unfavored kaon FFs are equal is used.
The charm contribution to the multiplicities is not taken into
account. For the value of $\chi^2/DOF$ corresponding to the best
fit to the data we obtain 137.8/77=1.79. A good description of the
COMPASS kaon data is achieved (for the quality of the fit see Fig.
1).
\begin{figure}[t!]
\begin{center}
  \includegraphics[height=.23\textheight]{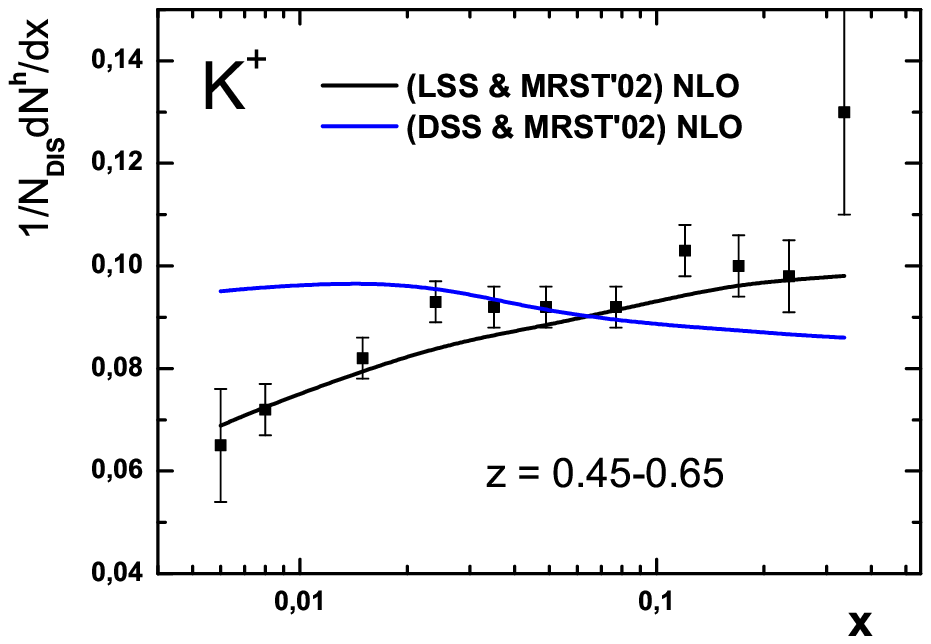}
\includegraphics[height=.23\textheight]{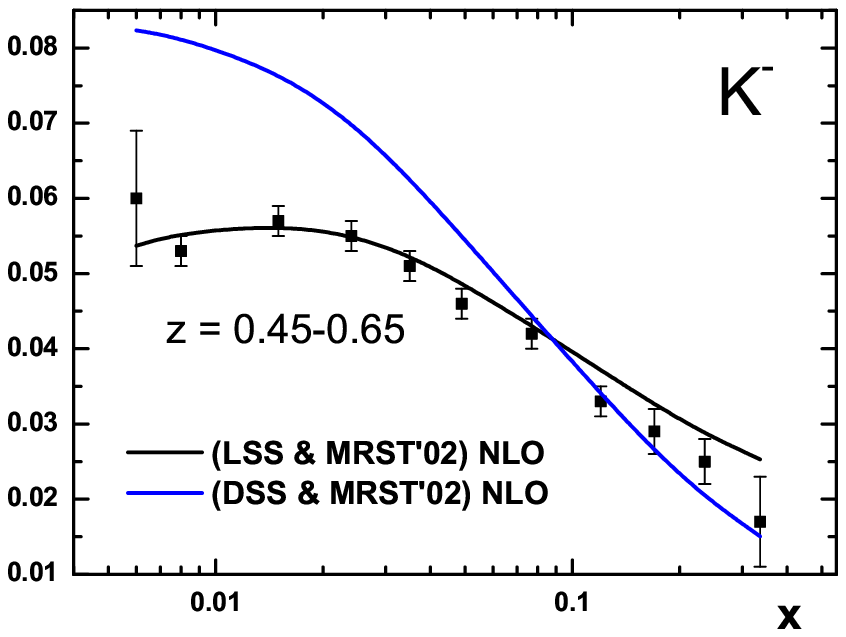}
\caption{\footnotesize Comparison between COMPASS kaon data for
$z_3$-bin and the best fit curves corresponding to the new FFs.
The curves corresponding to DSS FFs are also presented. In both
the cases the NLO MRST'02 set was used for the unpolarized PDFs.}
\label{Sidorov fig1}
\end{center}
\end{figure}
The new NLO kaon FFs are presented in Fig. 2.
\begin{figure}[h!]
\begin{center}
\includegraphics[height=.45\textheight]{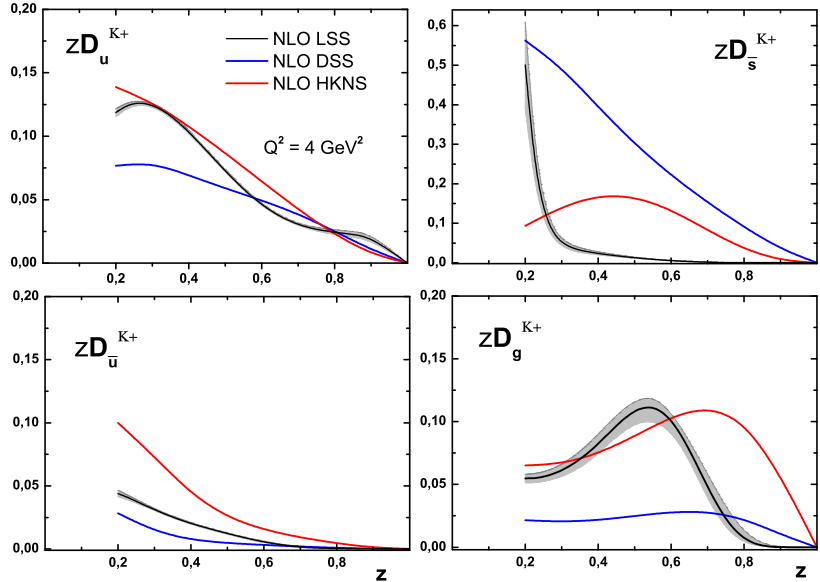}
\caption{\footnotesize Comparison between NLO LSS, DSS and HKNS
kaon FFs at $Q^2=4~GeV^2$. } \label{Sidorov fig2}
\end{center}
\end{figure}
As seen from Fig. 2 the new FFs are very different from those of
DSS and HKNS (2nd ref. in \cite{other_FFs}). This is especially
the case for the favored $D_{\bar s}^{K+}(z,Q^2)$ which changes
very rapidly between z=0.2 and z=0.3.

\begin{wrapfigure}[15]{R}{70mm}
  \centering 
  \vspace*{-5mm} 
  \includegraphics[width=65mm]{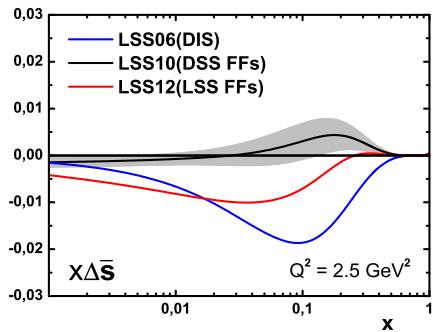}
  \caption{\footnotesize
Comparison between polarized strange quark densities obtained from
different kinds of NLO QCD analyses (see the text).}
\label{Sidorov_fig3}
\end{wrapfigure}
Using the new pion and kaon FFs we have performed a combined NLO
QCD analysis of the world polarized inclusive and semi-inclusive
DIS data in order to study their impact on the polarized parton
densities. Compared to the values of the polarized sea quark
densities obtained in our analysis \cite{LSS10} using the DSS FFs,
the changes are as follows: negligible for $\Delta \bar u(x)$,
visible for $\Delta \bar d(x)$, but still within the error band,
and {\it significant} for $\Delta \bar s(x)$. As seen from Fig. 3,
$x\Delta \bar s(x)$ is negative for any $x$ in the measured region
and {\it consistent} with that obtained from the pure DIS analysis
\cite{LSS07}. The {\it final} COMPASS and HERMES data on the
hadron multiplicities will be crucial in reliably determining the
fragmentation functions and polarized PDFs, as well as for the
resolution of the strange quark polarization puzzle.\\

This research was supported by the JINR-Bulgaria Collaborative
Grant and the RFBR Grants (Nrs 10-02-01259 and 11-01-00182).

\end{document}